\documentclass[twocolumn,showpacs,aps,pra,amsfonts,amsmath,amssymb,floatfix]
{revtex4}
\usepackage[dvips]{graphicx}

\newcommand{\affA}{%
     National Institute of Information and Communications Technology, \\
     4-2-1 Nukui-kitamachi, Koganei, Tokyo 184-8795, Japan \\
     CREST, Japan Science and Technology Agency, 
     1-9-9 Yaesu, Chuoh-ku, Tokyo 103-0028, Japan}

\begin{document}
\title{Discrimination of the binary coherent signal: 
Gaussian-operation limit and simple non-Gaussian near-optimal receivers
}
\date{\today}
\author{Masahiro Takeoka}
\author{Masahide Sasaki}
\address{\affA}%

\begin{abstract} 

We address the limit of the Gaussian operations and classical communication 
in the problem of quantum state discrimination. 
We show that the optimal Gaussian strategy for the discrimination 
of the binary phase shift keyed (BPSK) coherent signal 
is a simple homodyne detection. 
We also propose practical near-optimal quantum receivers 
that beat the BPSK homodyne limit in all areas of the signal power. 
Our scheme is simple and does not require realtime electrical feedback. 

\end{abstract} 
% 03.67.Hk Quantum communication  
% 42.50.-p Quantum optics 

\maketitle

%%%%%%%%%%%%%%%%%%%%%%%%%%%%%%%%%%%%%%%%%%%%%%%%%%%%%%%%%%%
\section{Introduction}
%%%%%%%%%%%%%%%%%%%%%%%%%%%%%%%%%%%%%%%%%%%%%%%%%%%%%%%%%%%

Discrimination of the binary phase shift keyed (BPSK) coherent states 
$\{ |\alpha\rangle, |$$-$$\alpha\rangle \}$ with the minimum error is 
one of the most fundamental issues in optical communication 
and quantum signal detection theory. 
Coherent communication theory has been developed based on 
semiclassical theory where these signals are detected 
by homodyne measurement. 
For the signals with equal prior probabilities, 
the average error probability is given by 
$P_{err} = {\rm erfc}[\sqrt{2|\alpha|}]/2$. 
This is often called the shot noise limit or the homodyne limit, 
falling short of the conventional error free criterion ($10^{-9}$) 
when $|\alpha|^2 < 10$.

It is, however, well known that the quantum optimal receiver 
can largely surpass the homodyne limit. 
The optimal measurement is mathematically given by 
a two-dimensional projection measurement and 
it attains the minimum error probability of 
$(1-\sqrt{1-e^{-4|\alpha|^2}})/2$ 
which is called the Helstrom bound \cite{QDET}. 
Kennedy proposed a simple near-optimal receiver 
using a coherent local oscillator (LO) and photon counting \cite{Kennedy73}. 
Its error rate is only twice larger than the Helstrom bound and is 
smaller than the homodyne limit when $|\alpha|^2 > 0.4$. 
Dolinar then extended this `Kennedy receiver' 
to the optimal one 
by introducing the adaptive electrical feedback 
which is enough faster than the optical signal pulse width \cite{Dolinar73} 
(see also \cite{QDET,Sasaki96,Geremia04,Takeoka05,Takeoka06}). 
Although the Dolinar's concept has been demonstrated recently \cite{Cook07}, 
it is still challenging to experimentally beat the homodyne limit 
with this approach because of its complicated system.

From a quantum mechanical point of view, 
homodyne measurement belongs to the class of Gaussian operations, 
i.e. described by up to the second order nonlinearity, 
while photon counting is the non-Gaussian one. 
The role of Gaussian operations in quantum information protocols 
\cite{Braunstein05} or quantum state estimation \cite{Hayashi05}
have been widely investigated. 
On the other hand, it has also been shown that some of the important protocols 
cannot be performed by only Gaussian operations and classical communication 
(GOCC) and inevitably requires non-Gaussian operations, e.g. 
quantum computing \cite{Bartlett02}, entanglement distillation of 
Gaussian states \cite{Eisert02,Fiurasek02,Giedke02}, and 
the optimal cloning of coherent states \cite{Cerf05}.

In this paper, we first show that 
the homodyne measurement is the best strategy to 
discriminate the binary coherent states within GOCC. 
To our knowledge, this is 
the first result addressing the Gaussian limit 
in quantum state discrimination scenario. 
In the second half of the paper, we propose novel non-Gaussian 
quantum receivers based on the Kennedy receiver, 
that beat the Gaussian limit for {\it any} $|\alpha|^2$. 
In particular, we point out that the amount of displacement 
in the Kennedy receiver is not optimal. 
Our schemes do not require realtime feedback and are simple and practical 
to experimentally overcome the homodyne limit with current technology.

%%%%%%%%%%%%%%%%%%%%%%%%%%%%%%%%%%%%%%%%%%%%%%%%%%%%%%%%%%%
\section{Discrimination 
%of the BPSK coherent signals 
via Gaussian operations and conditional dynamics}
%%%%%%%%%%%%%%%%%%%%%%%%%%%%%%%%%%%%%%%%%%%%%%%%%%%%%%%%%%%

In this section, we show that 
``the minimum error discrimination 
of a set of two coherent states $\{ |\alpha\rangle,|$$-$$\alpha\rangle \}$ 
with the prior probabilities $\{ p_+, p_- \}$ under GOCC 
is attained by the homodyne detection''.
For simplicity, $\alpha$ is assumed to be real. 
Gaussian operation is defined as the operation that maps 
Gaussian states to Gaussian states. 
%It is known that any Gaussian operation 
%(Gaussian completely positive (CP) map) in optical system 
%can be implemented by adding some ancillary system 
%prepared in Gaussian states, 
%Gaussian unitary operation on the whole system of the signal and ancilla 
%(which is implementable via linear optics and squeezing), 
%and then discarding and/or 
%performing homodyne measurements on 
%a part of the system. 
%%These Gaussian means are experimentally available. 
%We also include the conditional dynamics 
%via classical communication (CC), where 
%the partial measurement outcome from the system 
%is utilized to appropriately design 
%the next (quantum) operation applied to the rest of the system. 
%Characteristics of Gaussian operations with Gaussian input states 
%have been investigated 
%and 
%\cite{Fiurasek02,Giedke02}. 
%For example, any probabilistic (trace-decreasing) Gaussian operation 
%can be rewritten by a deterministic (trace-preserving) Gaussian operation 
%\cite{EFG02}. 
%For Gaussian input states, its property is well investigated and 
%it is known that 
%for any probabilistic (trace-decreasing) 
%Gaussian operation, 
%%when the input state is Gaussian, 
%there exists 
%a deterministic (trace-preserving) Gaussian operation 
%which has the same input-output property \cite{EFG02}. 
%It means that to perform a quantum protocol with Gaussian states and 
%operations, any classical communication (CC) 
%does not increase its performance. 
For Gaussian input states, properties of 
Gaussian operations have been well investigated 
\cite{Fiurasek02,Giedke02}. 
In our problem, however, although each signal state is given by 
a Gaussian state, 
the signal from the receiver's viewpoint is an ensemble of these states, 
$\hat{\rho}_i=p_+ |\alpha\rangle\langle\alpha| 
+ p_- |$$-$$\alpha\rangle\langle-\alpha|$, i.e. non-Gaussian. 
This is because 
the receiver does not know which state he or she is receiving.  
We therefore start by revisiting the measurement processes based on 
GOCC.

\subsection{Characterization of the measurements with GOCC}

It is known that any Gaussian operation 
(Gaussian completely positive (CP) map) in optical system 
can be implemented by adding an ancillary system 
prepared in Gaussian state, 
applying Gaussian unitary operation on the whole system 
(implementable via linear optics and squeezing), 
and then discarding and/or 
performing homodyne measurements on 
a part of the system \cite{Giedke02}. 
The CP map including measurements is not always trace-preserving 
and the output quantum state might be conditioned on the measurement outcome. 
When Gaussian operations are sequentially applied, 
the measurement outcomes (classical information) are sometimes useful to 
dynamically renew each step of quantum operations, 
which is called conditional dynamics. 
In the following, we characterize two types of measurements 
consisting of Gaussian operation with and without conditional dynamics.

\begin{figure}
\begin{center}
\includegraphics[width=1\linewidth]{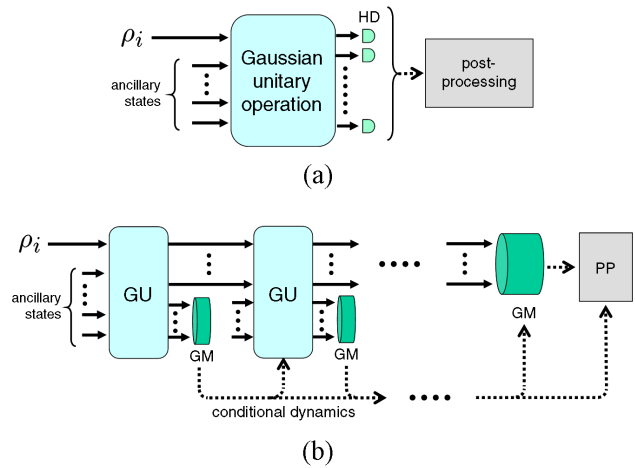}   %
\caption{\label{fig:GPOVM}
(Color online) 
Generic physical models of (a) Gaussian measurement 
and (b) GOCC-measurement. 
The solid and dotted lines represent quantum and 
classical signals, respectively. 
HD: homodyne detector, 
GU: Gaussian unitary operation, 
GM: Gaussian measurement, 
PP: post-processing.
}
\end{center}
\end{figure}

The first one is the measurement with 
only Gaussian quantum operation 
(without conditional dynamics). 
Here we call it a `Gaussian measurement'. 
%which we call a `Gaussian positive operator-valued measure (GPOVM)'.
A generic physical model of the Gaussian measurement is depicted 
in Fig.~\ref{fig:GPOVM}(a), consisting of 
Gaussian unitary operation, Gaussian ancillary states, 
and homodyne detectors. 
After the Gaussian measurement is finished, 
a set of measurement outcome is classically post-processed, 
in our case, to make the decision which signal is detected. 
Throughout this section, we consider only `noise-free' operations and 
measurements, that is 
we assume that 
ancillary states are always pure and 
the system is never discarded. 
It does not lose generality. 
Since to discard some of the modes means to lose their information, 
it is realized by measuring them via homodyne detectors 
and ignoring the outcomes,  
where the latter is included in classical post-processing. 
Mixed ancillary states are provided by first preparing 
entangled pure states and then discarding a part of them, 
therefore, can be prepared by pure states and post-processing. 
These imply the generality of the noise-free model. 
A Gaussian measurement detecting an $L$-mode quantum state 
is mathematically described by a positive operator-valued measure (POVM) 
$\{\hat{\Pi}_{\rm G}(\Gamma , \delta)\}_{\delta}$ where 
$\hat{\Pi}_{\rm G}(\Gamma , \delta) > 0$, 
$\int d \delta \, \hat{\Pi}_{\rm G}(\Gamma , \delta) = \hat{I}$, 
and $\hat{I}$ is an identity operator 
(see Appendix for its derivation from the physical model). 
The operator $\hat{\Pi}_{\rm G}(\Gamma , \delta)$ is a Gaussian operator, 
i.e. its characteristic function is described by 
$\chi(\omega) = \exp[-\frac{1}{4} \omega^T \Gamma \omega + 
i \delta^T \omega]$ 
where $\Gamma$ and $\delta$ are the $2L \times 2L$ covariance matrix 
and the $2L$-dimensional displacement vector, respectively, 
and $T$ is the transpose operation.

The second one consists of GOCC 
which we call a `GOCC-measurement'. 
As illustrated in Fig.~\ref{fig:GPOVM}(b), its generic model is 
described by sequential Gaussian operations with conditional dynamics 
via classical communication. 
Each step of Gaussian operation includes Gaussian operation 
and ancillary states, 
and a Gaussian measurement detecting a part of the system. 
%which is mathematically described by 
%a Gaussian trace-decreasing CP map \cite{Giedke02}. 
The measurement outcomes are applied to modify the following step of 
Gaussian operations in realtime, which is the conditional dynamics 
via classical communication. 
After the whole quantum measurement process is finished, 
all of the measurement outcomes is used for the classical 
post-processing.
The whole process except the post-processing
is described by a POVM 
$\{ \hat{\Pi}_{\rm GOCC}(\Gamma(x), \delta(x)) \}_x$ 
with a covariance matrix 
$\Gamma(x)$ and a displacement $\delta(x)$ where 
$\hat{\Pi}_{\rm GOCC}(\Gamma(x), \delta(x)) > 0$ and 
$\int d x \, \hat{\Pi}_{\rm GOCC}(\Gamma(x), \delta(x)) = \hat{I}$. 
The parameter $x$ indicates the pattern of which conditional dynamics 
is applied during the whole process. 
Again, throughout the section, we restrict the GOCC-measurements 
to be noise-free. 
%ancillary modes are initially prepared in pure states 
%and never discarded during the operations. 

Gaussian measurement is well characterized and easily applied 
to the optimization problem on the state discrimination. 
On the other hand, 
although GOCC-measurement is also well defined, 
it is not easy to handle its POVM directly. 
To prove the optimality of the homodyne limit under GOCC-measurements, 
therefore, we first show that the optimal Gaussian measurement without CC 
is a homodyne measurement. 
Then we discuss an important property of the conditional output 
from a Gaussian operation 
with an input of binary coherent state signals. 
Finally, we prove that even in the GOCC-measurement 
scenario, conditional dynamics is not useful and 
thus a simple homodyne measurement is optimal.

%%%%%%%%%%%%%%%%%%%%%%%%%%%%%%%%%%%%%%%%%%%%
%%%%%%%%%%%%%%%%%%%%%%%%%%%%%%%%%%%%%%%%%%%%

\subsection{Optimal Gaussian measurement}

A Gaussian measurement for the single-mode input state is 
described by 
$\{\hat{\Pi}(\Gamma_{\mathcal{M}} , D_{\mathcal{M}})\}_{D_{\mathcal{M}}}$ 
where $D_{\mathcal{M}}$ is a two-dimensional vector and 
%%%%%%%%%%%%%%%%%%%%%%%%%%%%%%
\begin{equation}
\label{eq:gamma_M}
\Gamma_{\mathcal{M}} = \left[ 
\begin{array}{cc}
c_- & s \\
s & c_+ 
\end{array}
\right] ,
\end{equation}
%%%%%%%%%%%%%%%%%%%%%%%%%%%%%%
where 
$c_\pm = \cosh (2r) \pm \sinh (2r) \cos\varphi$, 
$s = \sinh (2r) \sin\varphi$, 
and $r$ and $\varphi$ are the real parameters. 
The minimum error probability to discriminate 
$\{ |$$\pm$$\alpha\rangle, p_\pm \}$ by 
$\{\hat{\Pi}(\Gamma_\mathcal{M} , D_\mathcal{M})\}_{D_\mathcal{M}}$ 
with given $r$ and $\varphi$ 
can be calculated from the probability distribution to detect 
each signal 
$P_\pm (D_\mathcal{M}) = \langle\pm\alpha| 
\hat{\Pi}(\Gamma_\mathcal{M} , D_\mathcal{M}) |$$\pm$$\alpha\rangle$. 
Applying a conventional Bayesian decision strategy 
as a post-processing, 
we obtain 
%%%%%%%%%%%%%%%%%%%%%%%%%%%%%%
\begin{eqnarray}
\label{eq:Bayes_error_probability}
P_{e}^{(G)} & = & 
\frac{p_+}{2} {\rm erfc} \left[ e \sqrt{2}\alpha 
+ \frac{\ln (p_+/p_-)}{4 e \sqrt{2}\alpha } \right] 
\nonumber\\ && 
+ \frac{p_-}{2} {\rm erfc} \left[ e \sqrt{2}\alpha 
- \frac{\ln (p_+/p_-)}{4 e \sqrt{2}\alpha } \right] ,
\end{eqnarray}
%%%%%%%%%%%%%%%%%%%%%%%%%%%%%%
where 
%%%%%%%%%%%%%%%%%%%%%%%%%%%%%%
\begin{equation}
\label{eq:e}
e = \frac{1 + \cosh (2r) + \sinh (2r) \cos\varphi}{
2(1 + \cosh (2r))} . 
\end{equation}
%%%%%%%%%%%%%%%%%%%%%%%%%%%%%%
%and we have assumed that $\alpha$ is real 
%without loss generality. 
It is apparent that $P_{e}^{(G)}$ is minimum when 
$\varphi = 0$ and $r = \infty$, which implies that 
the homodyne detection with the phase $\varphi = 0$ 
is the optimal strategy within all possible Gaussian measurements.

\subsection{Conditional output states from a Gaussian operation}

As mentioned above, a GOCC-measurement consists of a sequence of 
Gaussian operations that include partial measurements. 
In this subsection, 
before discussing a whole GOCC-measurement process, 
we pick up one step of the sequence and 
address a useful property of the conditional output 
from a Gaussian operation with the binary coherent state inputs.

Let us consider the noise-free conditional Gaussian operation
which transforms a single-mode input to an $N$-mode output
where the output state is conditioned 
on a partial measurement outcome $d_\mathcal{M}$. 
Suppose an input state is $|\alpha\rangle$ or $|-\alpha\rangle$. 
We show that for any $d_\mathcal{M}$,  
the conditional output states of the inputs $|\pm\alpha\rangle$ 
can always be transformed to 
$|\alpha'_\pm\rangle\langle\alpha'_\pm| \otimes \hat{\rho}'_{\rm aux}$ 
by the same $d_\mathcal{M}$-independent deterministic Gaussian operation, 
where $\alpha'_\pm = \pm \alpha' + \bar{\alpha}'(d_\mathcal{M})$ 
and $\alpha'$ is independent of $d_\mathcal{M}$. 
We also show that, for a statistical ensemble input 
$\hat{\rho}_i = p_+|\alpha\rangle\langle\alpha| 
+ p_-|-\alpha\rangle\langle-\alpha|$, 
the conditional output is similarly transformed to be 
$\hat{\rho}'_i \otimes \hat{\rho}'_{\rm aux}$ where 
%%%%%%%%%%%%%%%%%%%%%%%%%%%%%%%
\begin{eqnarray}
\label{eq:rho_out}
\hat{\rho}'_i = p'_+ (d_\mathcal{M}) 
|\alpha'_+\rangle\langle\alpha'_+| 
+ p'_- (d_\mathcal{M}) 
|\alpha'_-\rangle\langle\alpha'_-| .
\end{eqnarray}
%%%%%%%%%%%%%%%%%%%%%%%%%%%%%%%

%Suppose an input state $\hat{\rho}_i = p_+|\alpha\rangle\langle\alpha| 
%+ p_-|-\alpha\rangle\langle-\alpha|$ is incident into 
%a 1-to-$N$ mode 
%Gaussian operation, which outputs 
%a partial measurement outcome $d_\mathcal{M}$ and 
%an $N$-mode conditional output state 
%$\hat{\rho}_{\rm out}$ depending on $d_\mathcal{M}$. 
%We show that for any $d_\mathcal{M}$,  
%$\hat{\rho}_{\rm out}$ can always be transformed 
%to $\hat{\rho}'_i \otimes \hat{\rho}'_{\rm aux}$ with 
%%%%%%%%%%%%%%%%%%%%%%%%%%%%%%%%
%\begin{eqnarray}
%\label{eq:rho_out}
%\hat{\rho}'_i = p'_+ (d_\mathcal{M}) 
%|\alpha'_+\rangle\langle\alpha'_+| 
%+ p'_- (d_\mathcal{M}) 
%|\alpha'_-\rangle\langle\alpha'_-| ,
%\end{eqnarray}
%%%%%%%%%%%%%%%%%%%%%%%%%%%%%%%%
%by $d_\mathcal{M}$-independent deterministic Gaussian operations, 
%where $\alpha'_\pm = \pm \alpha' + \bar{\alpha}'(d_\mathcal{M})$ 
%with $d_\mathcal{M}$-independent $\alpha'$. 
%It also transforms each of coherent states as 
%$|\pm\alpha\rangle \to |\alpha'_\pm\rangle\langle\alpha'_\pm|
%\otimes \hat{\rho}'_{\rm aux}$. 

Let $\hat{\rho}(\gamma,d)$ be a density matrix of a Gaussian state 
with a covariance matrix $\gamma$ and a displacement $d$. 
For example, the coherent states $|$$\pm$$\alpha\rangle$ are denoted as 
$\hat{\rho}(I_2, \pm d_\alpha)$ where 
$\pm d_\alpha = [\pm\sqrt{2}\alpha , 0 ]^T$ 
and $I_{2L}$ is a $2L \times 2L$ identity matrix.
%Let $\hat{\rho}(\gamma,d)$ be the density matrix of the Gaussian state 
%with the covariance matrix $\gamma$ and the displacement $d$. 
%In case of the coherent states $|$$\pm$$\alpha\rangle$, 
%these are given by $I_2$ and $\pm d_\alpha = [\pm\sqrt{2}\alpha , 0 ]^T$, 
%respectively, where $I_{2L}$ is a $2L \times 2L$ identity matrix.
%A generic implementation model of the generalized GPOVM is illustrated in 
%Fig.~\ref{fig:GPOVM}(a). 
%It consists of sequential Gaussian operations and CC. 
%Each Gaussian operation includes a Gaussian unitary operation, 
%auxiliary Gaussian states, and GPOVMs. 
%Here we temporally assume that the auxiliary states are pure states. 
%We first consider the two-step Gaussian measurement model 
%depicted in Fig.~\ref{fig:GPOVM}(b) and will generalize it later. 
%The two-step model consists of a Gaussian operation (first step), 
%GPOVM (second step), and CC. 
%The first step Gaussian operation may include a partial measurement 
%of the system and its measurement outcome is sent to the second step (CC)
%to optimally implement the second GPOVM. 
%Let us start from the first step. 
The conditional operation is described as follows. 
The initial single-mode state is interacted 
with $M-1$ Gaussian auxiliary states 
 ($M>N$) via a Gaussian unitary operation. 
Without loss of generality, we can set the auxiliary states 
to be $M-1$ vacua. 
At the covariance matrix level, Gaussian unitary operation is 
described by the matrix transformation via 
a symplectic matrix $S$ 
and an additional displacement $\bar{d}$. 
%that are desicribed by $2N \times 2N$ matrix, $S$, and 
%2$N$-dimensional vector, $\bar{d}$, respectively. 
These transform the coherent states 
$\hat{\rho}(I_2 , \pm d_\alpha)$ as 
%%%%%%%%%%%%%%%%%%%%%%%%%%%%%%%
\begin{eqnarray}
\label{eq:Gaussian_unitary_S}
I_2 & \to & S I_2 \oplus I_{2(M-1)} S^T 
\equiv \gamma, 
\\ 
\label{eq:Gaussian_unitary_d}
\pm d_\alpha & \to & 
S \left[ \pm \sqrt{2}\alpha , 0, \cdots , 0 \right]^T + \bar{d}
\equiv 
\pm d + \bar{d} ,
\end{eqnarray}
%%%%%%%%%%%%%%%%%%%%%%%%%%%%%%%
%where $S$ is the symplectic transformation corresponding to the Gaussian 
%unitary operation and 
where $S$ and $\gamma$ 
are $2M \times 2M$ matrices 
and $d$ and $\bar{d}$ are $2M$-dimensional vectors. 
The $N$-mode conditional output is obtained by performing 
an ($M-N$)-mode noise-free Gaussian measurement 
$\{ \hat{\Pi}(\gamma_\mathcal{M} , d_\mathcal{M}) \}_{d_\mathcal{M}}$. 
For convenience, we divide the system 
by the first $N$ modes and the remaining $M-N$ modes and call them 
the system A and B, respectively, as 
%%%%%%%%%%%%%%%%%%%%%%%%%%%%%%%
\begin{equation}
\label{eq:tilde_gamma_d}
\gamma = \left[
\begin{array}{cc}
A & C \\
C^T & B 
\end{array}
\right], \quad 
d 
= \left[
\begin{array}{c}
d^A \\
d^B
\end{array}
\right], \quad 
\bar{d} = \left[
\begin{array}{c}
\bar{d}^A \\
\bar{d}^B
\end{array}
\right]
.
\end{equation}
%%%%%%%%%%%%%%%%%%%%%%%%%%%%%%%
After performing the Gaussian measurement on the system $B$, 
each of the signals ($|\pm\alpha\rangle$) is transformed to 
an $N$-mode conditional output state as 
%%%%%%%%%%%%%%%%%%%%%%%%%%%%%%%
\begin{equation}
\label{eq:pm_alpha_out}
\hat{\rho}(I_2 , \pm d_\alpha) \to \hat{\rho} (\Gamma_{\rm out}, D_\pm) ,
\end{equation}
%%%%%%%%%%%%%%%%%%%%%%%%%%%%%%%
where \cite{Giedke02}
%%%%%%%%%%%%%%%%%%%%%%%%%%%%%%%
\begin{eqnarray}
\label{eq:gamma_out}
\Gamma_{\rm out} & = & A 
- C\frac{1}{B+\gamma_\mathcal{M}}C^T ,
\\ 
\label{eq:d_out}
D_\pm & = & 
\pm \left( d^A 
- C \frac{1}{B+\gamma_\mathcal{M}} 
d^B 
\right) 
\nonumber\\ &&
+ \bar{d}^A 
- C \frac{1}{B+\gamma_\mathcal{M}} 
\left( \bar{d}^B - d_\mathcal{M} \right) 
\nonumber\\ & \equiv & 
\pm D + \bar{D}_\mathcal{M} .
\end{eqnarray}
%%%%%%%%%%%%%%%%%%%%%%%%%%%%%%%
Note that each of $\hat{\rho} (\Gamma_{\rm out}, D_\pm)$ is a pure state 
since the operations are noise-free.

%%%%%%%%%%%%%%%%%%%%%%%%%%%%%%%%
%\begin{equation}
%\label{eq:rho_tilde_out}
%\tilde{\hat{\rho}}_{\rm out} = 
%p_+ \tilde{P}_+ (d_\mathcal{M}) \hat{\rho} (\tilde{\Gamma},\tilde{D}_+) 
%+ p_- \tilde{P}_- (d_\mathcal{M}) \hat{\rho} (\tilde{\Gamma},\tilde{D}_-) ,
%\end{equation}
%%%%%%%%%%%%%%%%%%%%%%%%%%%%%%%%
%where \cite{EFG02}
%%%%%%%%%%%%%%%%%%%%%%%%%%%%%%%%
%\begin{eqnarray}
%\label{eq:gamma_out}
%\tilde{\Gamma} & = & \tilde{A} 
%- \tilde{C}\frac{1}{\tilde{B}+\gamma_\mathcal{M}}\tilde{C}^T ,
%\\ 
%\label{eq:d_out}
%\tilde{D}_\pm & = & \tilde{d}^A_{\pm \, T} - 
%(\tilde{d}^B_\pm - d_\mathcal{M})^T 
%\frac{1}{\tilde{B}+\gamma_\mathcal{M}} \tilde{C}^T ,
%\\
%\label{eq:p_out}
%\tilde{P}_\pm (d_\mathcal{M}) & = & 
%\frac{1}{\sqrt{{\rm det}(\tilde{B} + \gamma_\mathcal{M})}} 
%\nonumber\\ && \times
%\exp\left[ -(\tilde{d}^B_\pm - d_\mathcal{M})^T 
%\frac{1}{\tilde{B}+\gamma_\mathcal{M}} 
%(\tilde{d}^B_\pm - d_\mathcal{M}) \right] . 
%\nonumber\\
%\end{eqnarray}
%%%%%%%%%%%%%%%%%%%%%%%%%%%%%%%%

%The $(N-M)$-mode output $\tilde{\hat{\rho}}_{\rm out}$ 
%is meausured by the second GPOVM. 
%The optimal structure of the second GPOVM may depend on
%the measurement outcome at the first step. 
%Before discussing this optimization problem, 
%we show the fact that 

Let us show that $\hat{\rho}(\Gamma_{\rm out},D_\pm)$ can be 
simultaneously transformed to 
$|\alpha'_\pm\rangle\langle\alpha'_\pm| \otimes \hat{\rho}'_{\rm aux}$ 
via Gaussian unitary operations. 
%Let us show that $\hat{\rho}_{\rm out}$ 
%can be transformed 
%to the form of $\hat{\rho}'_i \otimes \hat{\rho}'_{\rm aux}$. 
Since each of $\hat{\rho}(\Gamma_{\rm out}, D_\pm)$ is a pure state, 
there exists a symplectic transformation (i.e. 
Gaussian unitary operation) $S_D$ such that \cite{Holevo82}
%%%%%%%%%%%%%%%%%%%%%%%%%%%%%%%
\begin{equation}
\label{eq:symplectic_transformation}
\Gamma_{\rm out} \to S_D \Gamma_{\rm out} S_D^T 
= I_{2N} ,
\end{equation}
%%%%%%%%%%%%%%%%%%%%%%%%%%%%%%%
where the displacement is also transformed as 
%%%%%%%%%%%%%%%%%%%%%%%%%%%%%%%
\begin{eqnarray}
\label{eq:symplectic_transformation_d}
D_\pm & \to &
\pm S_D D + S_D \bar{D}_\mathcal{M}. 
\end{eqnarray}
%%%%%%%%%%%%%%%%%%%%%%%%%%%%%%%
Note that $S_D$ depends only on $\Gamma_{\rm out}$ 
and thus independent of $d_\mathcal{M}$. 
Let $S_D D \equiv [ d_1 , d_2 , \cdots , d_{2N} ]^T$ and 
$S_D \bar{D}_\mathcal{M} \equiv [ \bar{d}_1, \bar{d}_2, \cdots, 
\bar{d}_{2N} ]^T$.  
We can transform them to 
$[\pm d' , 0 , \cdots , 0]^T$ and 
$[\bar{d}'_1, \bar{d}'_2, \cdots, \bar{d}'_{2N} ]^T$, respectively,  
by some combination of linear optics (beamsplitters and phase shifters) 
where the covariance matrix $I_{2N}$ is kept to be invariant. 
Again parameters of the beamsplitters depend only on $\{ d_i \}_i$, 
and independent of $\{ \bar{d}_i \}_i$, i.e. 
free from $d_\mathcal{M}$. 
After these operations, the states are transformed to be 
the desired ones
%%%%%%%%%%%%%%%%%%%%%%%%%%%%%%%
\begin{eqnarray}
\label{eq:single-step_final_output_alpha}
\hat{\rho} (\Gamma_{\rm out},D_\pm) \to 
\hat{\rho} (I_2 , [\pm d' +\bar{d}'_1 , \bar{d}'_2]^T) 
\otimes \hat{\rho}'_{\rm aux} ,
\end{eqnarray}
%%%%%%%%%%%%%%%%%%%%%%%%%%%%%%%
where $\hat{\rho}'_{\rm aux}$ is a product of $N-1$ coherent states 
with the displacement 
$[ \bar{d}'_3, \bar{d}'_4, \cdots , \bar{d}'_{2N}]^T$. 
These are the desired ones. 

Finally we apply the above scenario onto 
the initial state of $\hat{\rho}_i = p_+ \hat{\rho}(I_2, +d_\alpha) 
+ p_- \hat{\rho}(I_2, -d_\alpha)$. 
Following the above procedures, its conditional output after 
the Gaussian operation is given by 
%%%%%%%%%%%%%%%%%%%%%%%%%%%%%%%
\begin{eqnarray}
\label{eq:rho_tilde_out}
\hat{\rho}_{\rm out} = 
p_+ P_+ (d_\mathcal{M}) \hat{\rho} (\Gamma_{\rm out},D_+) 
+ p_- P_- (d_\mathcal{M}) \hat{\rho} (\Gamma_{\rm out},D_-) ,
\nonumber\\
\end{eqnarray}
%%%%%%%%%%%%%%%%%%%%%%%%%%%%%%%
where 
%%%%%%%%%%%%%%%%%%%%%%%%%%%%%%%
\begin{eqnarray}
\label{eq:p_out}
P_\pm (d_\mathcal{M}) & = & 
\frac{1}{\sqrt{{\rm det}(B + \gamma_\mathcal{M})}} 
\exp\bigg[ 
-(\pm d^B + \bar{d}^B - d_\mathcal{M})^T 
\nonumber\\ && \times \left. 
\frac{1}{B+\gamma_\mathcal{M}} 
(\pm d^B + \bar{d}^B - d_\mathcal{M}) \right] , 
\end{eqnarray}
%%%%%%%%%%%%%%%%%%%%%%%%%%%%%%%
After the unitary operation of $S_D$ and 
appropriate linear operations, the state is transformed to be 
%%%%%%%%%%%%%%%%%%%%%%%%%%%%%%%
\begin{eqnarray}
\label{eq:single-step_final_output}
\hat{\rho}_{\rm out} \to \hat{\rho}'_i \otimes \hat{\rho}'_{\rm aux} ,
\end{eqnarray}
%%%%%%%%%%%%%%%%%%%%%%%%%%%%%%%
where 
%%%%%%%%%%%%%%%%%%%%%%%%%%%%%%%
\begin{eqnarray}
\label{eq:rho'_i}
\hat{\rho}'_i & = & 
p_+ P_+(d_\mathcal{M}) 
\hat{\rho} ( I_2, [d' + \bar{d}'_1 , \bar{d}'_2]^T ) 
\nonumber\\ && 
+
p_- P_-(d_\mathcal{M}) 
\hat{\rho} ( I_2, [-d' + \bar{d}'_1, \bar{d}'_2]^T ) .
\end{eqnarray}
%%%%%%%%%%%%%%%%%%%%%%%%%%%%%%%

%By tracing $\hat{\rho}'_{\rm aux}$ out, 
%the remaining state is to be a desired one, i.e. 
%$\hat{\rho}_{\rm out} = p'_+(d_\mathcal{M}) |\alpha'_+\rangle\langle\alpha'_+| 
%+ p'_-(d_\mathcal{M}) |\alpha'_-\rangle\langle\alpha'_-|$ 
%where $p'_\pm (d_\mathcal{M}) = p_\pm P_\pm (d_\mathcal{M} )$ 
%and $\alpha'_\pm = (\pm d' + \bar{d}'_1 + i \bar{d}'_2)/\sqrt{2}$, 
%and note that $d'$ is 
%independent on $d_\mathcal{M}$. 
%Note that the above transformation could be done without 
%knowing $d_\mathcal{M}$. 
%Also, if necessary, the transformation is reversible 
%in the sense that one can reproduce $\hat{\rho}'_{\rm aux}$ and 
%thus also $\tilde{\hat{\rho}}_{\rm out}$ 
%via a deterministic Gaussian operation 
%with the help of the information of $d_\mathcal{M}$. 

\subsection{Optimal GOCC-measurement}

Let us now turn to the state discrimination via GOCC-measurements. 
To specify the role of conditional dynamics, 
we first consider a simpler measurement scenario 
where a single conditional Gaussian operation 
and a Gaussian measurement are sequentially operated on the signal. 
Denote the partial measurement outcome at the former step 
as $d_\mathcal{M}$, which is informed to the latter measurement step 
to optimize the process of Gaussian measurement. 
After these GOCC processes, all measurement outcomes 
are classically post-processed.

Applying the result in the previous subsection to 
the initial state of $\hat{\rho}_i$, 
the conditional output from the first Gaussian operation 
can be transformed to corresponding $\hat{\rho}'_i$ described 
in Eq.~(\ref{eq:rho'_i}) 
via $d_\mathcal{M}$-independent deterministic Gaussian operations. 
Let this operation be a part of the second step Gaussian measurement 
(if necesssary, one can add $\hat{\rho}'_{\rm aux}$ as an ancilla). 
Then the remaining task in the measurement 
is to discriminate two coherent states 
$\{ |\alpha'_+\rangle, |\alpha'_-\rangle \}$ with 
the prior probabilities of $\{ p'_+(d_\mathcal{M}) , p'_-(d_\mathcal{M}) \}$. 
As already mentioned, the optimal Gaussian measurement is 
given by a simple homodyne detection. 
Its phase $\varphi$ is determined by geometric configuration between 
$\alpha'_+$ and $\alpha'_-$ and 
since $\alpha'_\pm = (\pm d' + \bar{d}'_1 + i\bar{d}'_2)/\sqrt{2}$ 
(see Eq.~(\ref{eq:rho'_i})), 
it is always given by $\varphi=0$ which is 
irrespective to the values of $\bar{d}'_1$, $\bar{d}'_2$ and thus also 
$d_\mathcal{M}$. 
An optimal strategy for the second step Gaussian measurement 
therefore consists of 
the transformation $\hat{\rho}_{\rm out} \to \hat{\rho}'_i$ 
and the homodyne detection where any parameters in those processes 
are independent of $d_\mathcal{M}$. 
It implies that the conditional dynamics is not necessary for 
to optimize the second step measurement \cite{comment1}. 
Consequently, the optimal whole process of these steps 
is described by a Gaussian measurement and thus, as already shown, is 
a homodyne measurement. 
Note that this statement is obtainable without specifying 
a concrete process of the first step Gaussian operation. 
An extension of the above scenario to the multi-step one is straightforward, 
which proves the optimality of the homodyne measurement within 
all possible GOCC-measurement.

\begin{figure}
\begin{center}
\includegraphics[width=1\linewidth]{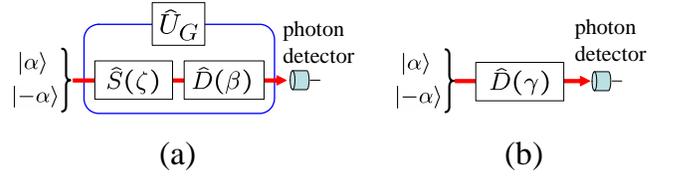}   %
\caption{\label{fig:schematic}
(Color online) 
Schematic of the near-optimal quantum receivers. 
(a) Type-I: Photon detector + optimal Gaussian unitary operation. 
(b) Type-II: Photon detector + optimal displacement. 
}
\end{center}
\end{figure}
%%%%%%%%%%%%%%%%%%%%%%%%%%%%%%%%%%%%%%%%%%%%

%%%%%%%%%%%%%%%%%%%%%%%%%%%%%%%%%%%%%%%%%%%%%%%%%%%%%%%%%%%
\section{Near-optimal quantum receiver by using a photon counter}
%%%%%%%%%%%%%%%%%%%%%%%%%%%%%%%%%%%%%%%%%%%%%%%%%%%%%%%%%%%

The homodyne limit (GOCC limit) stated in the previous section 
is overcome by adding a non-Gaussian measurement device. 
In this section, by extending the Kennedy receiver, 
we propose a simple near-optimal receiver 
where a photon counter, which is a typical non-Gaussian operation device, 
is added to the prior Gaussian operation. 
In what follows, we assume $p_+ = p_- = 1/2$ for simplicity.

In the Kennedy receiver, 
the BPSK signal $\{|\alpha\rangle, |-\alpha\rangle \}$ is 
shifted to $\{|2\alpha\rangle, |0\rangle\}$ by 
the displacement operation 
$\hat{D}(\alpha) = \exp(\alpha \hat{a}^\dagger - \alpha^* \hat{a})$, 
where $\hat{a}$ and $\hat{a}^\dagger$ are annihilation and 
creation operators, respectively, 
and then detected by an on/off type photon detector which discriminates 
zero or non-zero photons. 
It is well known that $\hat{D}(\alpha)$ can be realized 
by using a beamsplitter with the transmittance $\tau \to 1$ 
and the coherent LO $|\alpha/\sqrt{\tau}\rangle$. 
Here, we extend the Kennedy receiver and consider the setup 
depicted in Fig.~\ref{fig:schematic}(a), 
where the displacement $\hat{D}(\alpha)$ is replaced by 
a single-mode general Gaussian unitary operation $\hat{U}_G$. 
We will seek its optimal structure.

An on/off detector is described by the POVM 
$\{ \hat{\Pi}_{\rm off}, \hat{\Pi}_{\rm on} \}$ with 
%%%%%%%%%%%%%%%%%%%%%%%%%%%%%%
\begin{eqnarray}
\label{eq:on/off_POVM}
\hat{\Pi}_{\rm off} = 
e^{-\nu} \sum_{m=0}^\infty (1-\eta)^m |m \rangle\langle m| , 
\quad
\hat{\Pi}_{\rm on} = \hat{I} - \hat{\Pi}_{\rm off} , 
\end{eqnarray}
%%%%%%%%%%%%%%%%%%%%%%%%%%%%%%
where $|m\rangle$ is an $m$-photon state, 
$\eta$ is the quantum efficiency, and $\nu$ is the dark counts.
The Gaussian unitary operation $\hat{U}_G$ consists 
of phase shift, displacement, and squeezing, 
while one can omit the phase shift 
since the on/off detection is insensitive to 
the global phase. 
Then the average error probability is calculated from  
%%%%%%%%%%%%%%%%%%%%%%%%%%%%%%
\begin{equation}
\label{eq:Pe_average}
P_e = \frac{1}{2} \left( 
\langle\alpha| \hat{U}_G^\dagger \hat{\Pi}_{\rm off} 
\hat{U}_G |\alpha\rangle 
+ \langle -\alpha| \hat{U}_G^\dagger 
\hat{\Pi}_{\rm on} \hat{U}_G |-\alpha\rangle 
\right), 
\end{equation}
%%%%%%%%%%%%%%%%%%%%%%%%%%%%%%
where $\hat{U}_G = \hat{D}(\beta) \hat{S}(\zeta)$, 
$\hat{S}(\zeta) = \exp[ \frac{1}{2} (\zeta^* \hat{a}^2 
- \zeta \hat{a}^{\dagger\, 2}) ]$ is the squeezing operator, 
and $\zeta = r e^{i\varphi}$ is the complex squeezing parameter.

After some algebra, one can find that 
$r$, $\varphi$, and $\beta$ have the extreme at the same point, 
where $\varphi=0$, $\beta$ is real, and 
the optimal displacement $\beta_{\rm opt}$ and squeezing $r_{\rm opt}$, 
are given by $\beta$ and $r$ satisfying
%%%%%%%%%%%%%%%%%%%%%%%%%%%%%%
\begin{eqnarray}
\label{eq:optimal_r_beta}
\frac{8\eta\alpha\beta}{1-e^{4r}} & = & 
\left\{ \frac{4\eta (\alpha^2 + \beta^2)}{1-e^{4r}} 
- \frac{\eta + (2-\eta) e^{-2r}}{\eta + (2-\eta) e^{2r}}
\right\} 
\nonumber\\ && 
\times \tanh \left( 
\frac{4\eta\alpha\beta}{\eta+(2-\eta) e^{-2r}} \right) ,
\\
\label{eq:optimal_r_beta2}
\alpha & = & 
\beta \tanh \left( 
\frac{4\eta\alpha\beta}{\eta+(2-\eta) e^{-2r}} \right) ,
\end{eqnarray}
%%%%%%%%%%%%%%%%%%%%%%%%%%%%%%
simultaneously. 
The optimized average error probability is then given by 
%%%%%%%%%%%%%%%%%%%%%%%%%%%%%%
%\begin{widetext}
\begin{eqnarray}
\label{eq:Pe_average_beta_opt}
P_e^{DS} & = & \frac{1}{2} - 
\frac{2 e^{-\nu}}{\sqrt{(\eta+(2-\eta)e^{2r_{\rm opt}})
(\eta+(2-\eta)e^{-2r_{\rm opt}})}} 
\nonumber\\ && \times 
\exp\left[ 
-\frac{2 \eta (\alpha^2 + \beta_{\rm opt}^2)
}{\eta + (2-\eta)e^{-2r_{\rm opt}}} 
\right]
\nonumber\\ && \times 
\sinh\left[ \frac{4\eta\alpha\beta_{\rm opt}
}{\eta + (2-\eta)e^{-2r_{\rm opt}}} \right] .
\end{eqnarray}
%\end{widetext}
%%%%%%%%%%%%%%%%%%%%%%%%%%%%%%
In the following, we call it as the Type-I receiver. 
It should be noted that if one can use an arbitrarily higher order 
nonlinear unitary operation instead of $\hat{U}_G$, 
it is able to achieve the Helstrom bound rigorously \cite{Sasaki96}.

\begin{figure}
\begin{center}
\includegraphics[width=0.8\linewidth]{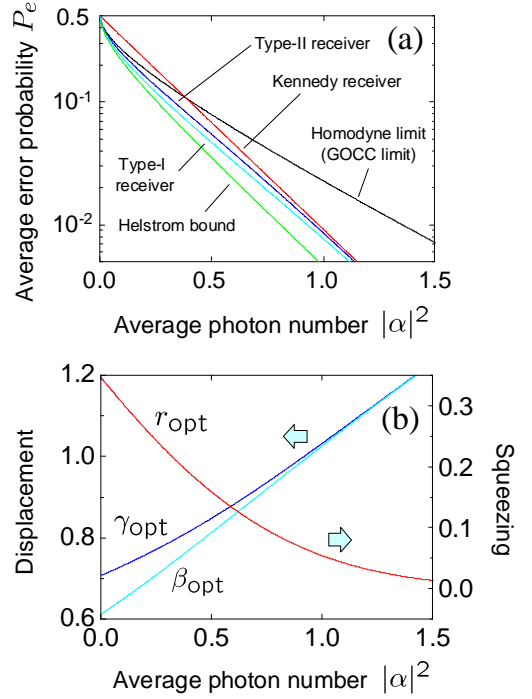}   %
\caption{\label{fig:Pe_comparison}
(Color online) 
(a) Average error probabilities for 
the Type-I, Type-II, and Kennedy receivers, the homodyne limit, 
and the Helstrom bound. 
(b) The optimal displacements and squeezing
for the Type-I and II receivers. 
}
\end{center}
\end{figure}

On the other hand, if one is restricted to use only linear unitary 
operation, that is the displacement $\hat{D}(\gamma)$, 
the conditions in Eqs.~(\ref{eq:optimal_r_beta}) and 
(\ref{eq:optimal_r_beta2}) are simplified as 
%%%%%%%%%%%%%%%%%%%%%%%%%%%%%%
\begin{equation}
\label{eq:optimal_gamma}
\alpha = \gamma \tanh (2\eta\alpha\gamma) ,
\end{equation}
%%%%%%%%%%%%%%%%%%%%%%%%%%%%%%
The schematic is shown in Fig.~\ref{fig:schematic}(b) and 
we call it the Type-II receiver. 
Its average error probability is given by 
%%%%%%%%%%%%%%%%%%%%%%%%%%%%%%
\begin{equation}
\label{eq:Pe_average_gamma_opt}
P_e^{D} = \frac{1}{2} - e^{-\nu-\eta(\alpha^2 + \gamma_{\rm opt}^2)} 
\sinh \left(2 \eta\alpha\gamma_{\rm opt}\right) ,
\end{equation}
%%%%%%%%%%%%%%%%%%%%%%%%%%%%%%
where $\gamma_{\rm opt}$ is the $\gamma$ satisfying 
Eq.~(\ref{eq:optimal_gamma}). 
Note that its physical setup is the same as that of the Kennedy receiver. 
However, we stress that $\gamma_{\rm opt} \ne \alpha$ in general 
and thus the conventional Kennedy receiver is easily improved by 
using $\hat{D}(\gamma_{\rm opt})$ instead of $\hat{D}(\alpha)$. 
Figure~\ref{fig:Pe_comparison}(a) plots the average error probabilities 
for the Type-I, Type-II, and Kennedy receivers, the homodyne limit, 
and the Helstrom bound while $r_{\rm opt}$, $\beta_{\rm opt}$, and 
$\gamma_{\rm opt}$ are shown in Fig.~\ref{fig:Pe_comparison}(b). 
It is shown that the error probabilities for both the Type-I and II receivers 
are better than the homodyne limit for any $|\alpha|^2$.

%%%%%%%%%%%%%%%%%%%%%%%%%%%%%%%%%%%%%%%%%%%%
%%%%%%%%%%%%%%%%%%%%%%%%%%%%%%%%%%%%%%%%%%%%

\begin{figure}
\begin{center}
\includegraphics[width=0.75\linewidth]{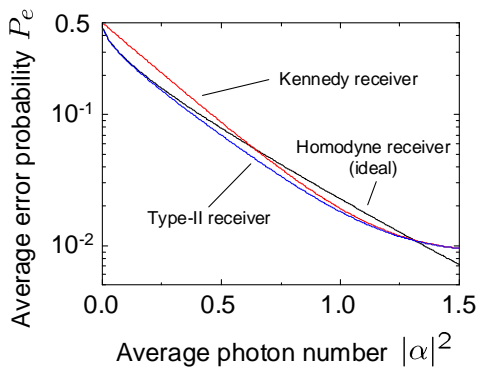}   %
\caption{\label{fig:Pe_imperfection}
(Color online) 
Average error probabilities for the ideal homodyne receiver 
and the Type-II and Kennedy receivers 
with practical imperfections, $\tau=0.99$, $\eta=0.9$, $\nu=10^{-3}$, 
and $\xi=0.995$.
}
\end{center}
\end{figure}

Let us finally discuss the practical perspective of these 
non-Gaussian receivers, particularly, the Type-II receiver. 
The superiority of the Type-II receiver rather than the Kennedy receiver 
in $|\alpha|^2\le1$ is significant to beat the homodyne limit 
in realistic experiments. 
It is known that the Kennedy-type receiver is not robust against 
thermal noise or dark counts \cite{Vilnrotter84}. 
Moreover, even without environmental noises, 
the mode mismatch between the signal and LO causes additional dark counts. 
As mentioned above, the displacement $\hat D (\beta)$ is realized by 
interfering the signal with the coherent state LO 
$|\beta/\sqrt{1-\tau}\rangle$ via the beamsplitter of 
the transmittance $\tau$. 
The effect of mode mismatch can be characterized by introducing 
the mode match factor $\xi$ ($0\le\xi\le1$) representing the overlap 
between the signal and LO pulse areas. 
Since these two pulses are in a coherent state, the average intensity 
of the signal field after the interference is simply given by 
%%%%%%%%%%%%%%%%%%%%%%%%%%%%%%
\begin{equation}
\label{eq:Intensity}
I = (1-\xi) \left( \tau |\alpha|^2 + |\beta|^2 \right) 
+ \xi \left| \pm \sqrt{\tau}\alpha + \beta \right|^2 .
\end{equation}
%%%%%%%%%%%%%%%%%%%%%%%%%%%%%%
Due to its Poissonian photon number distribution, 
the average discrimination error including 
$\tau$ and $\xi$ at the on/off detector is described as 
%%%%%%%%%%%%%%%%%%%%%%%%%%%%%%
\begin{equation}
\label{eq:Pe_average_gamma_opt_modemismatch}
\tilde P_e^{D} = \frac{1}{2} - 
e^{-\nu-\eta(\tau\alpha^2 + \tilde\gamma_{\rm opt}^2)} 
\sinh \left(2 \eta\xi\sqrt{\tau}\alpha\tilde\gamma_{\rm opt}\right) ,
\end{equation}
%%%%%%%%%%%%%%%%%%%%%%%%%%%%%%
where $\tilde\gamma_{\rm opt}$ fulfills the optimality condition 
%%%%%%%%%%%%%%%%%%%%%%%%%%%%%%
\begin{equation}
\label{eq:optimal_gamma_modemismatch}
\xi\sqrt{\tau}\alpha = \tilde\gamma_{\rm opt}  
\tanh (2\eta\xi\alpha\tilde\gamma_{\rm opt}) .
\end{equation}
%%%%%%%%%%%%%%%%%%%%%%%%%%%%%%
An example of the average error probabilities including the imperfections
is shown in Fig.~\ref{fig:Pe_imperfection} which clearly shows 
the advantage of our proposed receiver would be crucial 
to experimentally observe the gain of the non-Gaussian measurement 
beyond the homodyne limit. 
Although the requirement for $\eta$ in the weaker signal is still high, 
recent experimental progress in this field is rather promising 
\cite{Rosenberg05}.

%%%%%%%%%%%%%%%%%%%%%%%%%%%%%%%%%%%%%%%%%%%%
%%%%%%%%%%%%%%%%%%%%%%%%%%%%%%%%%%%%%%%%%%%%

%%%%%%%%%%%%%%%%%%%%%%%%%%%%%%%%%%%%%%%%%%%%%%%%%%%%%%%%%%%
\section{Conclusions}
%%%%%%%%%%%%%%%%%%%%%%%%%%%%%%%%%%%%%%%%%%%%%%%%%%%%%%%%%%%

In this paper, we have addressed the discrimination of the BPSK signals 
and proved that the homodyne limit is the minimum error probability 
attainable via Gaussian operations and classical communication. 
This is the first clarification of the limit of 
Gaussian operation in the state discrimination problem. 
Although it is shown for the binary coherent states 
that any conditional dynamics due to CC is not effective, 
we note that this would not be the case 
for the discrimination of more than two signals. 
Related to this topic, an increase of the mutual information by 
the adaptive homodyne strategies has been numerically observed 
\cite{Bargatin05}. 
For further investigation into this direction, 
more useful formulation of the GOCC-measurement would be necessary.

We have also proposed the near-optimal quantum receivers 
for the BPSK coherent signals, that are based on a photon detector 
and Gaussian operations. 
Our schemes are simple and do not require realtime electrical feedback 
although their error probabilities are better than the homodyne limit 
for any signal photon number region. 
Because of the recent experimental progress of 
high efficiency photon detectors \cite{Rosenberg05} 
and universal squeezing operations \cite{Yoshikawa07}, 
we believe that now it would be feasible in near future 
to beat the homodyne limit in digital optical communication experiments.

{\it Note added:} Proof-of-principle experiment of the Type-II receiver 
is recently demonstrated \cite{Wittmann08}.

\acknowledgements

We would like to thank stimulating and encouraging discussions 
with U.~L.~Andersen, M.~Ban, R.~Filip, L.~Mista, and C.~Wittmann. 
We also acknowledge valuable comments from the referee. 
This work was supported by a MEXT Grant-in-Aid for 
Young Scientists (B) 19740253.

\appendix*
\section{Generic model of the GPOVM}

Here we show that the POVM of 
the physical model illustrated in Fig.~\ref{fig:GPOVM}(a) 
is always described by $\{ \hat{\Pi}_G (\Gamma,\delta) \}_\delta$ 
introduced in Sec.~IIA. 
Let $\hat{\rho}_{\rm sig}^A$ and $\hat{\rho}_{\rm aux}^B$ 
be an $N_A$-mode input state and 
an $N_B$-mode ancillary state, respectively. 
The probability distribution of the measurement is given by 
%%%%%%%%%%%%%%%%%%%%%%%%%%%%%%
\begin{eqnarray}
\label{eq:GPOVM_1}
p(d_{\rm HD}) & = & 
{\rm Tr}_{AB} \left[ \left(
\hat{U}_S^{AB} \hat{\rho}_{\rm sig}^A \otimes 
\hat{\rho}_{\rm aux}^B \hat{U}_S^{AB\,\dagger}
\right) 
\hat{\Pi}_{\rm HD}^{AB}(\Gamma_{\rm HD}, d_{\rm HD}) \right]
\nonumber\\ & = & 
{\rm Tr}_A \left[ \hat{\rho}_{\rm sig}^A 
{\rm Tr}_B \left[ 
\hat{\rho}_{\rm aux}^B \hat{U}_S^{AB\,\dagger} 
\hat{\Pi}_{\rm HD}^{AB} (\Gamma_{\rm HD}, d_{\rm HD}) 
\hat{U}_S^{AB} \right] \right] ,
\nonumber\\
\end{eqnarray}
%%%%%%%%%%%%%%%%%%%%%%%%%%%%%%
where $\hat{U}_S^{AB}$ is an ($N_A+N_B$)-mode Gaussian unitary operation 
and $\{ \hat{\Pi}_{\rm HD}^{AB}(\Gamma_{\rm HD}, d_{\rm HD}) \}$ represents 
$N_A+N_B$ homodyne detectors with the measurement outcomes 
denoted by $d_{\rm HD}$. 
Note that homodyne detection is a Gaussian measurement 
(projection onto an infinitely squeezed states) and thus 
characterized by the covariance matrix. 
In Eq.~(\ref{eq:GPOVM_1}), $\Gamma_{\rm HD}$ is 
a $2(N_A+N_B) \times 2(N_A+N_B)$ diagonal matrix and 
%%%%%%%%%%%%%%%%%%%%%%%%%%%%%%
\begin{eqnarray}
\label{eq:gamma_HD}
\Gamma_{\rm HD} = {\rm diag}[ e^{-2r} , e^{2r} , e^{-2r} , e^{2r} , 
\cdots , e^{2r} ] , 
\end{eqnarray}
%%%%%%%%%%%%%%%%%%%%%%%%%%%%%%
with $r \to \infty$. 
%The measurement outcome $d_{\rm HD}$ is a $2(N_A+N_B)$ vector. 

Equation (\ref{eq:GPOVM_1}) implies that the POVM of 
the Gaussian measurement model is derived from a set of operators 
$\{
{\rm Tr}_B [ \hat{\rho}_{\rm aux}^B \hat{U}_S^{AB\,\dagger} 
\hat{\Pi}_{\rm HD}^{AB} (\Gamma_{\rm HD}, d_{\rm HD}) 
\hat{U}_S^{AB} ] \}_{d_{\rm HD}}$. 
Let us describe it by covariance matrices and displacements. 
Denoting the symplectic transformation corresponding to 
$\hat{U}_S^{AB}$ by $S$, the unitary transformation 
$\hat{U}_S^{AB\,\dagger} 
\hat{\Pi}_{\rm HD}^{AB} (\Gamma_{\rm HD}, d_{\rm HD}) 
\hat{U}_S^{AB}$ is described by 
%%%%%%%%%%%%%%%%%%%%%%%%%%%%%%
\begin{eqnarray}
\label{eq:S_G-meas}
\Gamma_{\rm HD} & \to & S^T \Gamma_{\rm HD} S 
\equiv \Gamma_S , 
\\
d_{\rm HD} & \to & S d_{\rm HD} 
\equiv d_S . 
\end{eqnarray}
%%%%%%%%%%%%%%%%%%%%%%%%%%%%%%
Then after tracing out the ancillary system $B$, 
we find that the above POVM is given by 
an $N_A$-mode Gaussian operator $\hat{\Pi}_G (\Gamma, \delta)$ 
with 
%%%%%%%%%%%%%%%%%%%%%%%%%%%%%%
\begin{eqnarray}
\label{eq:gamma_delta_G-meas}
\Gamma & = & \Gamma_A - 
\Gamma_C \frac{1}{\Gamma_{\rm aux} + \Gamma_B} \Gamma_C^T ,
\\
\delta & = & d_A - 
\Gamma_C \frac{1}{\Gamma_{\rm aux} + \Gamma_B} d_B ,
\end{eqnarray}
%%%%%%%%%%%%%%%%%%%%%%%%%%%%%%
where we have denoted 
%%%%%%%%%%%%%%%%%%%%%%%%%%%%%%
\begin{eqnarray}
\label{eq:gamma_delta_G-meas_2}
\Gamma_S & = & \left[
\begin{array}{cc}
\Gamma_A & \Gamma_C \\
\Gamma_C^T & \Gamma_B 
\end{array}
\right] ,
\\
d_S & = & \left[ 
\begin{array}{c}
d_A \\ d_B 
\end{array}
\right] .
\end{eqnarray}
%%%%%%%%%%%%%%%%%%%%%%%%%%%%%%


\begin{references}

\bibitem{QDET}
   C.~W.~Helstrom, 
   {\it Quantum Detection and Estimation Theory} 
   (Academic Press, New York, 1976).

\bibitem{Kennedy73}
   R.~S.~Kennedy, 
   Research Laboratory of Electronics, MIT, 
   Quarterly Progress Report No.~108, 1973 (unpublished), p.~219.

\bibitem{Dolinar73}
   S.~Dolinar, 
   Research Laboratory of Electronics, MIT, 
   Quarterly Progress Report No.~111, 1973 (unpublished), p.~115.

\bibitem{Sasaki96}
  M.~Sasaki and O.~Hirota, 
  Phys.\ Rev.\ A\,\textbf{54}, 2728 (1996).

\bibitem{Geremia04}
  J.~M.~Geremia, 
  Phys.\ Rev.\ A\,\textbf{70}, 062303 (2004).  

\bibitem{Takeoka05}
  M.~Takeoka, M.~Sasaki, P.~van Loock, and N.~L\"{u}tkenhaus, 
  Phys.\ Rev.\ A\,\textbf{71}, 022318 (2005). 

\bibitem{Takeoka06}
  M.~Takeoka, M.~Sasaki, and N.~L\"{u}tkenhaus, 
  Phys.\ Rev.\ Lett.\,\textbf{97}, 040502 (2006).

\bibitem{Cook07}
   R.~L.~Cook, P.~J.~Martin, and J.~M.~Geremia, 
   Nature {\bf 446}, 774 (2007). 

\bibitem{Braunstein05}
   S.~L.~Braunstein and P.~van Loock, 
   Rev.\ Mod.\ Phys.\, \textbf{77}, 513 (2005). 

\bibitem{Hayashi05}
   {\it Asymptotic theory of quantum statistical inference: 
selected papers}
   (edited by M.~Hayashi, World Scientific Publishing, New York, 1976).

\bibitem{Bartlett02}
  S.~D.~Bartlett, B.~C.~Sanders, S.~L.~Braunstein, 
  and K.~Nemoto, 
  Phys.\ Rev.\ Lett.\,\textbf{88}, 097904 (2002).

\bibitem{Eisert02}
  J.~Eisert, S.~Scheel, and M.~B.~Plenio, 
  Phys.\ Rev.\ Lett.\,\textbf{89}, 137903 (2002). 

\bibitem{Fiurasek02}
  J.~Fiurasek, {\it ibid},\,\textbf{89}, 137904 (2002). 

\bibitem{Giedke02}
  G.~Giedke and J.~I.~Cirac, 
  Phys.\ Rev.\ A\,\textbf{66}, 032316 (2002). 

\bibitem{Cerf05}
  N.~J.~Cerf, O.~Kr\"{u}ger, P.~Navez, R.~F.~Werner, and M.~M.~Wolf, 
  Phys.\ Rev.\ Lett.\,\textbf{95}, 070501 (2005). 

%\bibitem{Demoen77}
%  B.~Demoen, P.~Vanheuverzwijn, and A.~Verbeure, 
%  Lett.\ Math.\ Phys.\,\textbf{2}, 161 (1977). 
%
%\bibitem{Eisert02-2}
%  J.~Eisert and M.~B.~Plenio, 
%  Phys.\ Rev.\ Lett.\,\textbf{89}, 097901 (2002). 


%\bibitem{comment1}
%It is obvious that the POVM consisting of only Gaussian operations 
%and Gaussian ancillae is described by a set of Gaussian operators. 
%On the other hand, as mentioned in the text, any GPOVM is physically 
%implemented by a heterodyne-type measurement. 

%\bibitem{comment2}
%One may trace out some of the ancillary modes, which results 
%higher rank POVM elements. 
%However, the trace out operation can be understood 
%such that one makes homodyne measurements and ignores their outcome. 
%Obviously we can exclude such possibility in our optimization process. 



\bibitem{Jamiolkowski72}
  A.~Jamio{\l}kowski, 
  Rep.\ Math.\ Phys.\,\textbf{3}, 275 (1972). 

\bibitem{Holevo82}
  A.~S.~Holevo, 
  {\it Probabilistic and Statistical Aspects of Quantum Theory} 
  (North-Holland, Amsterdam, 1982) Chap.~5. 

\bibitem{comment1}
Note that here we mention the dynamical optimization of only 
`quantum' operations.  Generally, the measurement outcomes at both steps are 
necessary for the optimal classical post-processing. 

\bibitem{Vilnrotter84}
  V.~A.~Vilnrotter and E.~R.~Rodemich, 
  IEEE\ Trans.\ Inf.\ Theory\,\textbf{30}, 446 (1984). 

\bibitem{Bargatin05}
  I.~Bargatin, 
  Phys.\ Rev.\ A\,\textbf{72}, 022316 (2005). 

\bibitem{Rosenberg05}
  D.~Rosenberg, A~E.~Lita, A.~J.~Miller, and S.-W. Nam, 
  Phys.\ Rev.\ A\,\textbf{71}, 061803(R) (2005). 

\bibitem{Yoshikawa07}
  J.~Yoshikawa, T.~Hayashi, T.~Akiyama, N.~Takei, A.~Huck, 
  U.~L.~Andersen, and A.~Furusawa, 
  Phys.\ Rev.\ A\,\textbf{76}, 060301(R) (2007).

\bibitem{Wittmann08}
  C.~Wittmann, M.~Takeoka, K.~N.~Cassemiro, M.~Sasaki, G.~Leuchs, 
  and U.~L.~Andersen, Submitted (2008).


\end{references}
\end{document}